\documentclass{PoS}
\usepackage{slashed}
\usepackage{amsmath}
\newcommand{\bs}[1]{{\bf #1}}

\title{Mass anomalous dimension of SU$(2)$ using the spectral density method}

\ShortTitle{Mass anomalous dimension of SU$(2)$ using the spectral density method}

\author{\speaker{Joni M. Suorsa}\\
        Helsinki Institute of Physics and Department of Physics, University of Helsinki\\
        E-mail: \email{joni.suorsa@helsinki.fi}}

\author{Viljami Leino\\ 
		Helsinki Institute of Physics and Department
		of Physics, University of Helsinki\\ E-mail:
		\email{viljami.leino@helsinki.fi}}

\author{Jarno Rantaharju\\
		CP$^{ \bf 3}${-Origins}, IFK \& IMADA, University of Southern
		Denmark\\ E-mail: \email{rantaharju@cp3.dias.sdu.dk}}

\author{Teemu Rantalaiho\\ 
		Helsinki Institute of Physics and Department
		of Physics, University of Helsinki\\ E-mail:
		\email{teemu.rantalaiho@helsinki.fi}}

\author{Kari Rummukainen\\
		Helsinki Institute of Physics and Department
		of Physics, University of Helsinki\\ E-mail:
		\email{kari.rummukainen@helsinki.fi}}

\author{Kimmo Tuominen \\
		Helsinki Institute of Physics and Department
		of Physics, University of Helsinki\\ E-mail:
		\email{kimmo.i.tuominen@helsinki.fi}}
\author{Sara Tähtinen \\
		Helsinki Institute of Physics and Department
		of Physics, University of Helsinki\\ E-mail:
		\email{sara.tahtinen@helsinki.fi}}

\abstract{SU$(2)$ with $N_f=6$ and $N_f = 8$ are believed to have an infrared conformal fixed point. We use the spectral density method cross referenced with the mass step scaling method to evaluate the coupling constant dependence of the mass anomalous dimension for massless HEX smeared, clover improved Wilson fermions with Schrödinger functional boundary conditions.}

\FullConference{34th annual International Symposium on Lattice Field Theory\\
		24-30 July 2016\\
		University of Southampton, UK}

\begin{document}

\section{Introduction}
Non-Abelian gauge theories with infrared-conformality have been considered as viable models for physics beyond the Standard Model. In these models, the anomalous dimension $\gamma_m$ of the fermion operator $\overline{\psi}\psi$ is a quantity of specific interest.
The scaling of the spectral density of the massless Dirac operator is
governed by the mass anomalous dimension \cite{deldebbio}, and 
while the explicit calculation of the eigenvalue distribution is prohibitively costly, recently developed stochastic methods \cite{luscher} have made it possible to determine the mass anomalous dimension from the scaling of the mode number of the Dirac operator \cite{patella}.

The theories which we are studying are SU$(2)$ with $N_f = 6$ and $8$ fermions in the fundamental representation. While $N_f = 8$ seems to be well within the conformal window \cite{viljami1, Frandsen:2010ej}, the situation with $N_f = 6$ has been unclear \cite{Karavirta:2011zg, Bursa:2010xn, Hayakawa:2013maa, Appelquist:2013pqa}. However, in our recent study \cite{viljami2} we have observed clear evidence of a fixed point.


The mode number of the Dirac operator is known to follow a scaling behaviour
 of
\begin{equation}\label{modenumber1}
\nu(\Lambda) \equiv 2\int_0^{\sqrt{\Lambda^2 - m^2}} \rho(\lambda)d\lambda \simeq \nu_0(m) +  C\left[\Lambda^2 - m^2\right]^{2/(1+\gamma_*)}
\end{equation} 
in some energy range between the infrared and the ultraviolet in the vicinity of a fixed point.
Here $\rho(\lambda)$ is the eigenvalue density of the Dirac operator, $\gamma_*$ is the mass anomalous dimension $\gamma_m$ at the fixed point, $\nu_0(m)$ is an additive constant, $C$ is a dimensionless constant that is a combination of renormalisation factors, and $m$ is the quark mass. The energy range where the power law behaviour of Eq. \ref{modenumber1} 
 holds is not 
known beforehand, and needs to be determined by observing the quality of the fit in a given range.


The mass anomalous dimension can also be obtained by using the Schr\"odinger functional mass step scaling method \cite{stepscaling}, and in what follows we will compare results obtained using both of these methods.

\begin{figure}
\begin{center}
\includegraphics[width=0.69\textwidth]{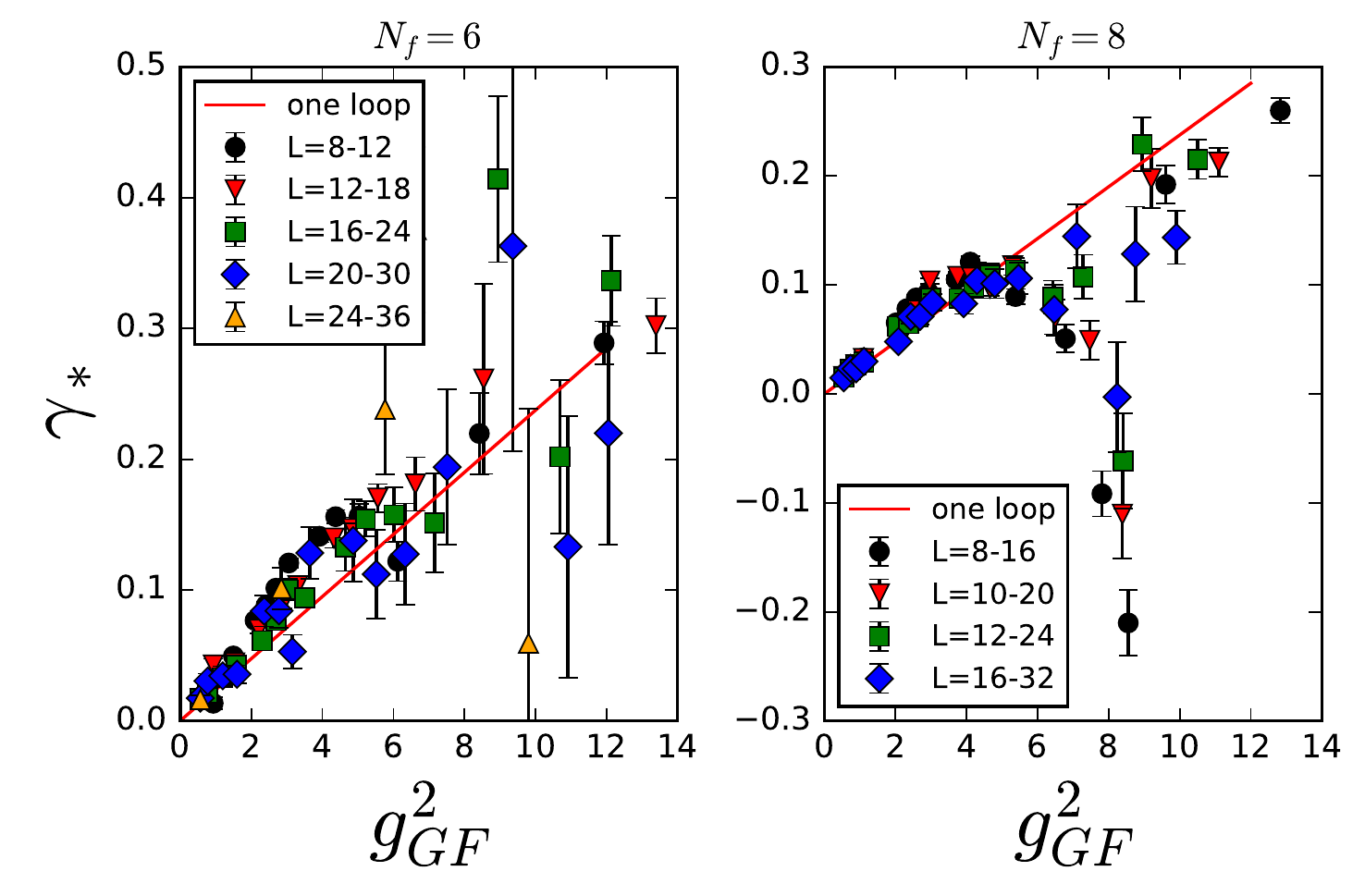}
\label{viljamiplotti}
\end{center}
\caption{The mass anomalous dimension as a function of the gradient flow coupling constant obtained using the mass step scaling function. The different symbols correspond to different lattice size pairings. For $N_f = 6 $ the fixed point is at $g_{GF}^2\sim 14.5$ \cite{viljami2}, and for $N_f = 8$ at $g_{GF}^2\sim 6$ \cite{viljami1}. The results for larger couplings become unstable.}
\end{figure}

\section{Mass step scaling}

We simulate SU$(2)$ with $N_f=6$ and $N_f = 8$ using HEX smeared \cite{hex}, clover improved \cite{clover} Wilson fermions
using the same parameters as for the evaluation of the running coupling using the gradient flow (GF) method \cite{viljami1, viljami2, sf}.
In order to study the massless case we tune the hopping parameter to $\kappa =\kappa_c$ for which $m_{PCAC}\sim 0$. For $N_f = 6 $ we simulate the theory at eight different values of $\beta$ corresponding to measured gauge couplings from $g_{GF}^2 = 0.56$ to $g_{GF}^2 = 14.24$ on a $V=24^4$ lattice, and for $N_f = 8 $ we use eight different values of $\beta$ corresponding to couplings from $g_{GF}^2 = 0.55$ to $g_{GF}^2 = 9.49$ on a $V=32^4$ lattice. 

For the evaluation of the mass anomalous dimension using the step scaling method, we use the Schr\"odinger functional boundary conditions:
\begin{equation}
 U_i(\mathbf{x},t=0) = U_i(\mathbf{x},t=L) = \mathbf{1}.
\end{equation}
The mass anomalous dimension $\gamma_m$ is measured
from the running of the pseudoscalar density renormalization constant
\cite{stepscaling,dellamorte}
\begin{align}
Z_P(L) = \frac{\sqrt{3 f_1} }{f_P(L/2)},
\label{Zp}
\end{align}
where
\begin{align}
 f_P(t) &= \frac{-a^6}{3L^6}\sum_{\bs y,\bs z}
 \langle P^a(\bs x, t)
 \,\bar\zeta(\bs y)\gamma_5\frac12\sigma^a\zeta(\bs z)\rangle, \\
 f_1 &= \frac{-a^{12}}{3 L^{12}} \sum_{\bs u,\bs v, \bs y, \bs z}
 \langle\bar \zeta'(\bs u)\gamma_5\frac12 \sigma^a \zeta'(\bs v)\,
  \bar\zeta(\bs y)\gamma_5\frac12 \sigma^a\zeta(\bs z)\rangle.
\end{align}

Here $P^a(x) = \overline{\psi}(x)\gamma_5 \frac{1}{2}\sigma^a \psi(x)$, and $\zeta$ and $\zeta'$ are boundary quark sources at $t=0$ and $t=L$ respectively.
The mass step scaling function is then defined as \cite{stepscaling}
\begin{align}
\Sigma_P(u,s,L/a) &=
   \left. \frac {Z_P(g_0,sL/a)}{Z_P(g_0,L/a)} \right |_{g_{GF}^2(g_0,L/a)=u}
   \label{Sigmap}\\
\sigma_P(u,s) &= \lim_{a/L\rightarrow 0} \Sigma_P(u,s,L/a).
\end{align}
For $N_f = 6$ we choose $s = 3/2$ and for $N_f = 8$ $s = 2$, and find the continuum step scaling function $\sigma_P$ by
measuring $\Sigma_P$ at $L/a=8$, $12$, $16$, $20, $ $24$ and $L/a=8$, $10$, $12$, $16$ for $N_f = 6$ and 8 respectively.
The step scaling function can then be used to obtain the mass anomalous dimension \cite{dellamorte} by
\begin{align}
  \gamma_*(u) = -\frac{\log \sigma_P(u,s)}{\log s }.
\label{eq:gammastar}
\end{align}

Our preliminary results using the mass step scaling method are shown in Fig. \ref{viljamiplotti}. The method gives results comparable to one loop perturbation theory predictions at small gauge coupling $g_{GF}^2$, but becomes unstable at large coupling as
the theory flows toward the fixed point at $g_{GF}^2 \sim 14.5$ for $N_f = 6$ \cite{viljami2} and at $g_{GF}^2 \sim 6$ for $N_f = 8$ \cite{viljami1}.

\section{Spectral density method}
We calculate the mode number per unit volume of Eq. \ref{modenumber1} by using 
\begin{equation}
\nu(\Lambda) =\lim_{V\to \infty}  \frac{1}{V}\left< \textnormal{tr }\mathbb{P}(\Lambda)\right>,
\end{equation}
where the operator $\mathbb{P}(\Lambda)$ projects from the full eigenspace of $M = m^2 - \slashed{D}^2$ to the eigenspace of eigenvalues smaller than $\Lambda^2$. The trace is evaluated stochastically \cite{luscher}. 

We use the lattices obtained from the step scaling analysis, and use between 12 to 20 well separated configurations for each value of the gauge coupling. We calculate the mode number for 90 values of $\Lambda^2$ ranging from $10^{-3}$ to $0.3$ for $N_f = 6$ and from $10^{-4}$ to $0.3$ for $N_f = 8$. The difference in the lower limits comes from the smaller lattice size of $N_f = 6$, for which the mode number reaches zero at larger $\Lambda$.

The two constants $\nu_0(m)$ and $m^2 = (Z_A m_{PCAC})^2$ in Eq. \ref{modenumber1} are expected to be negligible since we are studying the massless Dirac operator and the additive constant $\nu_0(m)$ is related to the part of the spectrum that is sensitive to the effects of the nonzero mass. In our analysis we used
\begin{equation}\label{modenumber2}
\nu(\Lambda) \simeq  C\Lambda^{4/(1+\gamma_*)}
\end{equation}
as the function we fit the calculated mode number data to, and checked that the error relative to  Eq. \ref{modenumber1} was $\mathcal{O}(10^{-3})$.

The fit range was determined by varying the lower and the upper limit of the fit range and observing the stability and the quality of the fit. As a cross reference we compared the value of $\gamma_*$ obtained using the spectral density method for small couplings to the value obtained using the step scaling method in order to further assess wether the chosen fit range was good or not. 

\begin{figure}[t]
\begin{center}
\includegraphics[width=0.69\textwidth]{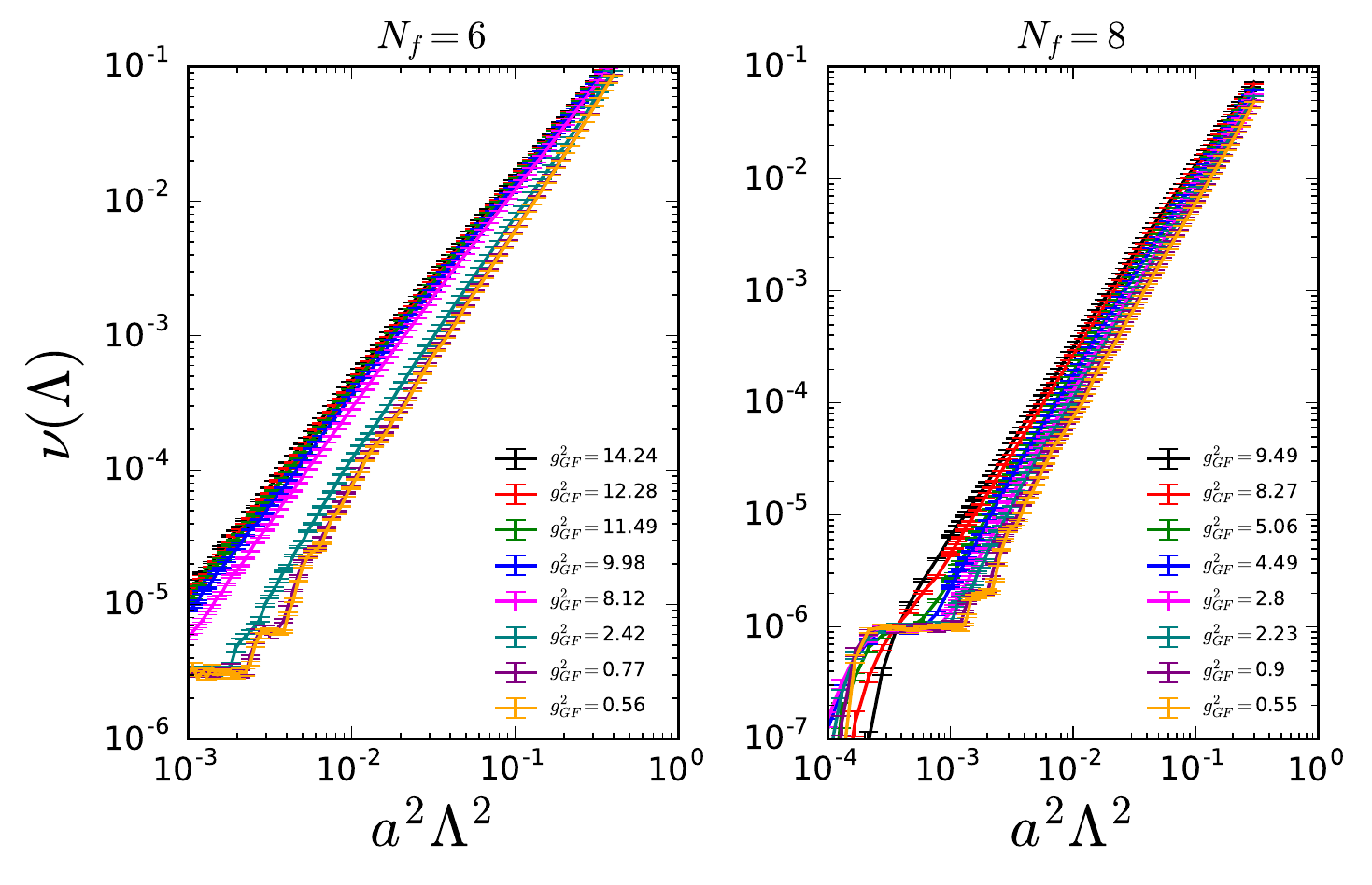}
\caption{The mode number calculated for different gauge couplings for $N_f = 6$ on a $V=24^4$ lattice, and for $N_f = 8$ on a $V = 32^4$ lattice.}
\label{all_beta}
\end{center}
\end{figure}

In Fig. \ref{all_beta} we present the mode number data we have calculated for both $N_f$. It is apparent that at small couplings the low eigenvalues appear in discrete energies which manifests in the step-like structure of the mode number curve. This behaviour should vanish when the volume goes to infinity.

\begin{figure}[t]
\begin{center}
\includegraphics[width=0.69\textwidth]{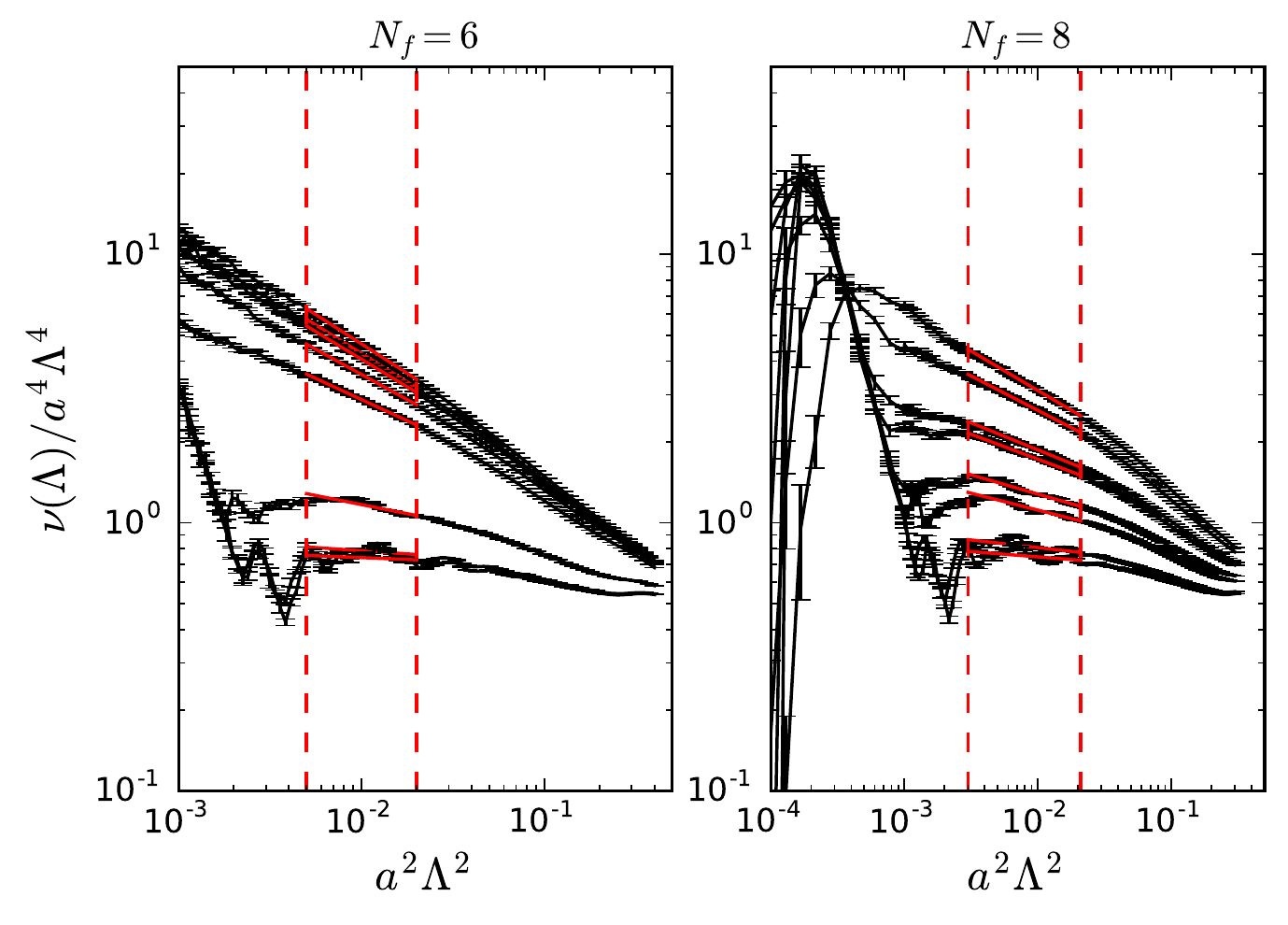}
\caption{The mode number divided by $a^4\Lambda^4$ as a function of $a^2\Lambda^2$. The dashed red lines indicate the chosen fit range and the red solid lines the fit function. The fit ranges were varied around these chosen regions. The curves are in a descending gauge coupling order.}
\label{fit_range}
\end{center}
\end{figure}

In Fig. \ref{fit_range} we plot the mode number divided by the fourth power of the eigenvalue scale with the chosen fit range and the fit function of Eq. \ref{modenumber2} shown overlaid in red. According to Eq. \ref{modenumber2} in the proximity of the fixed point the infrared behaviour should be approximately linear on a log-log plot and precisely linear exactly at the fixed point in the absence of lattice artefacts. Indeed we observe this behaviour at strongest couplings for both $N_f= 6 $ and 8. We use the same fit range for the weak coupling as for the large coupling to illustrate the evolution of the mass anomalous dimension, even though the power law is clearly not evident.

\begin{figure}[t]
\begin{center}
\includegraphics[width=0.69\textwidth]{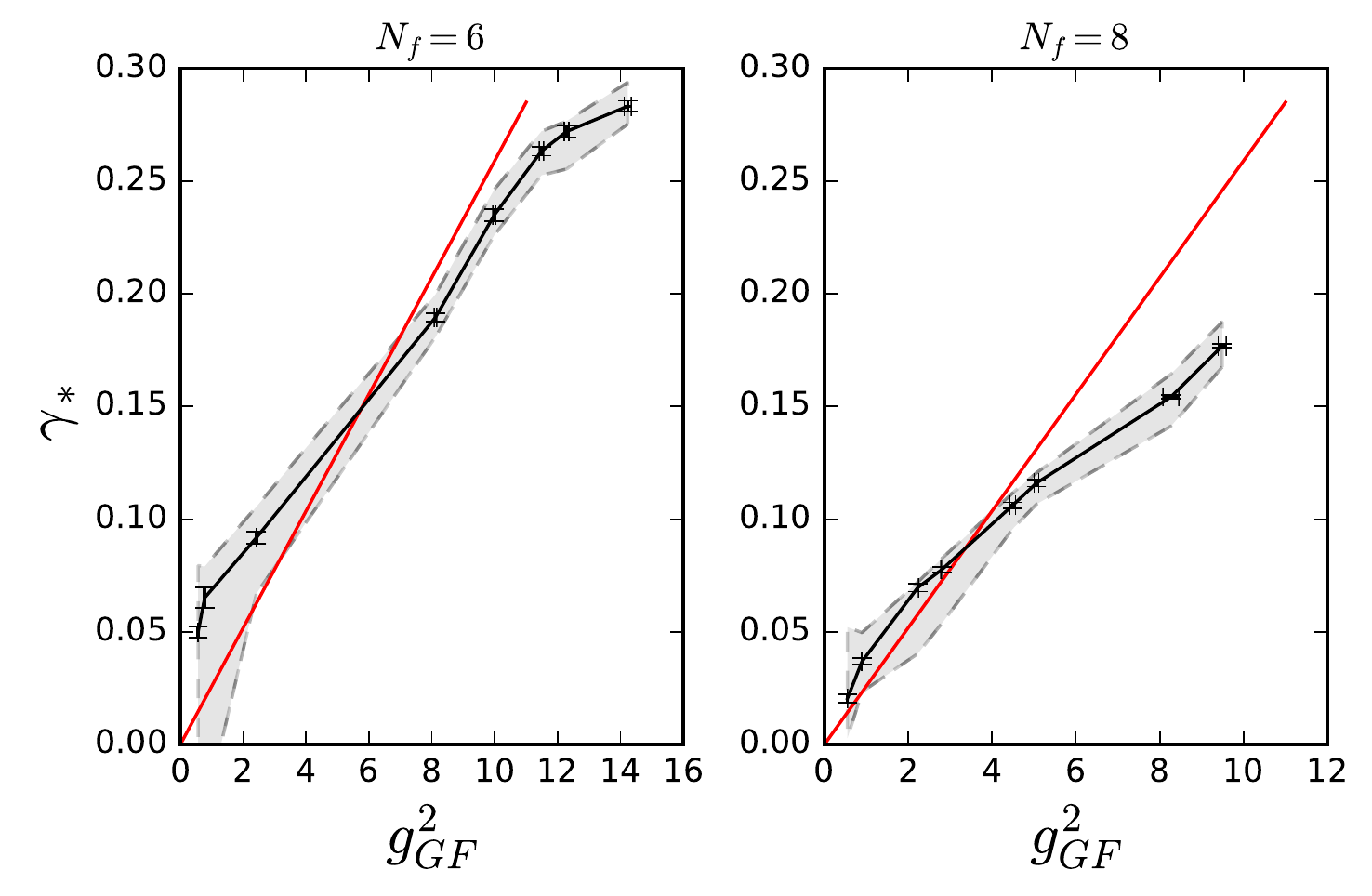}
\caption{The value of $\gamma_*$ obtained by fitting Eq. \protect\ref{modenumber2} to the data in Fig. \protect\ref{all_beta} is shown with black points and the one loop perturbative result with a red line. The shaded regions are estimates for reasonable ranges of values obtainable using the method, and were obtained by varying the fit range shown in Fig. \protect\ref{fit_range} slightly.}
\label{result}
\end{center}
\end{figure}

Our main results for the spectral density method are shown in Fig. \ref{result} where we plot the mass anomalous dimension $\gamma_*$ obtained from fitting Eq. \ref{modenumber2} to the data as a function of the gauge coupling $g_{GF}^2$. 


In order to quantify the uncertainty in choosing the fit range, we have varied the fit range around a chosen range which produced reasonable results, and this is represented as the shaded band around the curve. The largest uncertainty is with the smallest gauge couplings since the data is not smooth, and a slight change in the fit range changes the angle of the fit line dramatically. The larger couplings near the fixed point are not as sensitive to the fit range variation, which can be seen from the narrowing error bands in Fig. \ref{result} as one goes from small coupling to larger coupling.

While the results obtained using the mass step scaling method shown in Fig. \ref{viljamiplotti} show good agreement with the perturbative line, the spectral density method suffers from large errors for small couplings. For larger couplings near the fixed point the spectral density method shows consistent behaviour whereas the mass step scaling method seemed to break down.

\section{Conclusions}


We have determined the mass anomalous dimension of SU(2) gauge theory with six and eight Dirac fermions in the fundamental representation of the gauge group using the spectral density method and the mass step scaling method. We have demonstrated that the spectral density method gives results compatible with perturbation theory and the nonperturbative mass step scaling method at weak coupling.
With the spectral density method our estimate for the fixed point mass anomalous dimension is  $\gamma_* \sim 0.275$ for $N_f = 6$ and $\gamma_* \sim 0.125$ for $N_f = 8$. The precise error analysis of the error range remains to be completed.


A major source of uncertainty that is not easily quantifiable is the choice of the fit range where Eq. \ref{modenumber2} (or Eq. \ref{modenumber1}) is used to describe the data.
  For larger couplings near the fixed point the spectral density method works well and we observe behaviour that seems to be a genuine nonperturbative feature and not an artefact due to fit uncertainties. Thus, the spectral density method in conjunction with the step scaling method give access to the mass anomalous dimension reliably from small coupling to couplings near or at the fixed point.

\section*{Aknowledgements}
J.M.S. is supported by the Jenny and Antti Wihuri foundation. K.R., V.L., and K.T. are supported by the Academy of Finland
grants 267842, 134018, and 267286. J.R. is supported by the Danish
National Research Foundation grant number DNRF:90. T.R. and S.T. are supported by the Magnus Ehrnrooth foundation. The simulations were performed at the Finnish IT Center for Science (CSC) in Espoo, Finland. Parts of the simulation program have been derived from the MILC lattice simulation program \cite{milc}.

\end{document}